\documentclass[preprint2]{aastex}
\usepackage{graphicx}
\usepackage{color}

\begin{document}

\title{On a spectral pattern of the Von-Neumann probes}


\author{Osmanov Z.}
\affil{School of Physics, Free University of Tbilisi, 0183, Tbilisi,
Georgia}

\begin{abstract}
In this paper we have extended our previous work \citep{probe1} and considered spectral characteristics of interstellar non-relativistic Type-II and Type-III Von-Neumann probes. It has been shown that by means of the proton capture, required for replication, strong bremstrahlung emission will be generated. We have found that for both types of civilizations the probes might be visible mainly in the infrared spectral band, but as it has been found the probes might be visible in the ultraviolet as well. For both cases it was shown that differential power has dips for equally spaced frequencies, which might be a significant fingerprint to identify such exotic interstellar objects.
\end{abstract}


\section{Introduction}


In the previous paper (henceforth paper-I) we have considered self-replicating Von-Neumann extraterrestrial probes \citep{probe1}. In this work we considered self replicating extraterrestrial robots and estimated their lengthscales. We found that for making a process of colonization of space very efficient, the probes must be very small. Then, in a molecular cloud the typical timescale of replication might be several years, which might lead to potentially detectable very intense radiation. In Paper-I we estimated orders of magnitude of radiation frequency, bolometric luminosity and its increment, but did not study the spectral characteristics of emission, the study of which is a purpose of the present work. The idea of self-reproducing probes has been proposed in the last century by \cite{neum}. In particular, he studied the possibility of machines with artificial intelligence capable of self-reproduction and the problem has been examined in in the context of information theory and corresponding thermodynamics. This idea also has been revisited in the context of the search for extraterrestrial intelligence (SETI) \citep{GS}. After the discovery of "Oumuamua"'s unusual behaviour \citep{oum1}, the interest to the study of extraterrestrial probes has been revived, despite the natural explanation of the aforementioned peculiarity \citep{oum2}.

In general, the study of techno-signatures of advanced extraterrestrial civilizations has been actively discussed since a brilliant idea of \cite{dyson}, who proposed to search for Type-II \citep{kardashev} civilizations by searching for infrared spheres (Dyson spheres - DS) built around stars. It is worth noting that according to Kardashev's classification Type-I is a technological civilization which uses the total energy coming from the host Solar-type star to the Earth-like planet, Type-II is an advanced alien society harnessing the total energy emitted by the star and Type-III civilization is capable to use the total energy of the host galaxy. The interest to techno-signatures has revived after the discovery of a strange behaviour of the Tabby's star's (Name in the catalog KIC8462852) flux, characterised by unusual high dips \citep{kic846}. The problem is so complex that Dysonian SETI should be revisited and widened \citep{cirkovic1}. Interesting attempts have been presented in a series of papers \citep{wd,paper1,paper2,paper3,paper5,haliki,gaia,lacki}. 

In this context a special interest deserves our previous work (Paper-I) where we have studied the Von-Neumann interstellar Type-II probes\footnote{In paper-1 a) in Eq. (2) for the probe's size and density there should be 0.01$\mu$m and $1cm^{-3}$ and b) the optimal size is $10^{-6}$cm instead of $7\times 10^{-6}cm$.}. The main aim of the paper was a) to show high efficiency of small sized probes in colonising the space and b) estimate very general characteristics of the emission pattern. In the paper-I we have examined the probes with three possible velocities: $0.01c, 0.02c, 0.1c$ ($c$ is the speed of light) and it was shown that number of the probes increases drastically by the factor of $10^{32}$ during $650$yrs ($\upsilon = 0.01c$), $320$yrs ($\upsilon = 0.02c$), $60$yrs ($\upsilon = 0.1c$). Which are quite realistic values because they are small compared to a timescale ($3000$yrs) required for Type-I civilization to reach the level of Type-II \citep{dyson}. We have found that the probes might be seen from radio to the infrared spectral band and estimated the luminosity increment time-scales which lie in the interval $(1; 10) yrs$ and the corresponding achieved luminosities might be enormous, at least of the order of $10^{16-20}erg/s$.

The aim of the present paper is certain extension of  paper-I. In particular, we are going to study the spectral characteristics of the probes studying energy emitted by the unit of time per unit of frequency for Type-II and Type-III extraterrestrial probes. 

The paper is organized in the following way: in Sec. 2, we coonsider the general tools and methods, applying them to Type-II and Type-III civilizations and obtain corresponding results and in Sec. 3 we outline the summary of the  results.

\begin{figure}
  \centering {\includegraphics[width=7cm]{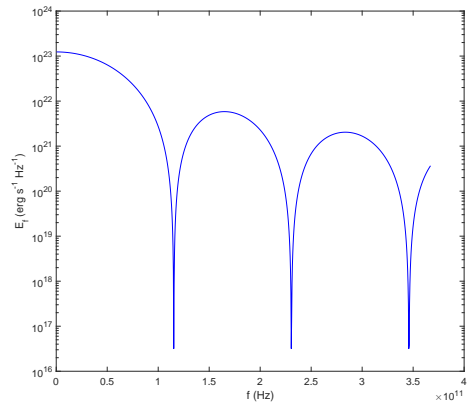}}
  \caption{The behaviour of $E_f$ versus frequency. The set of parameters is: $\beta = 0.01$, $N_0 = 100$, $\xi = 0.1$, $m_0 = m_p$ ($m_p$ is the proton mass), $\rho = 0.4g\;cm^{-3}$, $n = 10^4cm^{-3}$, $D = 2$ pc.}\label{fig1}
\end{figure}

\begin{figure}
  \centering {\includegraphics[width=7cm]{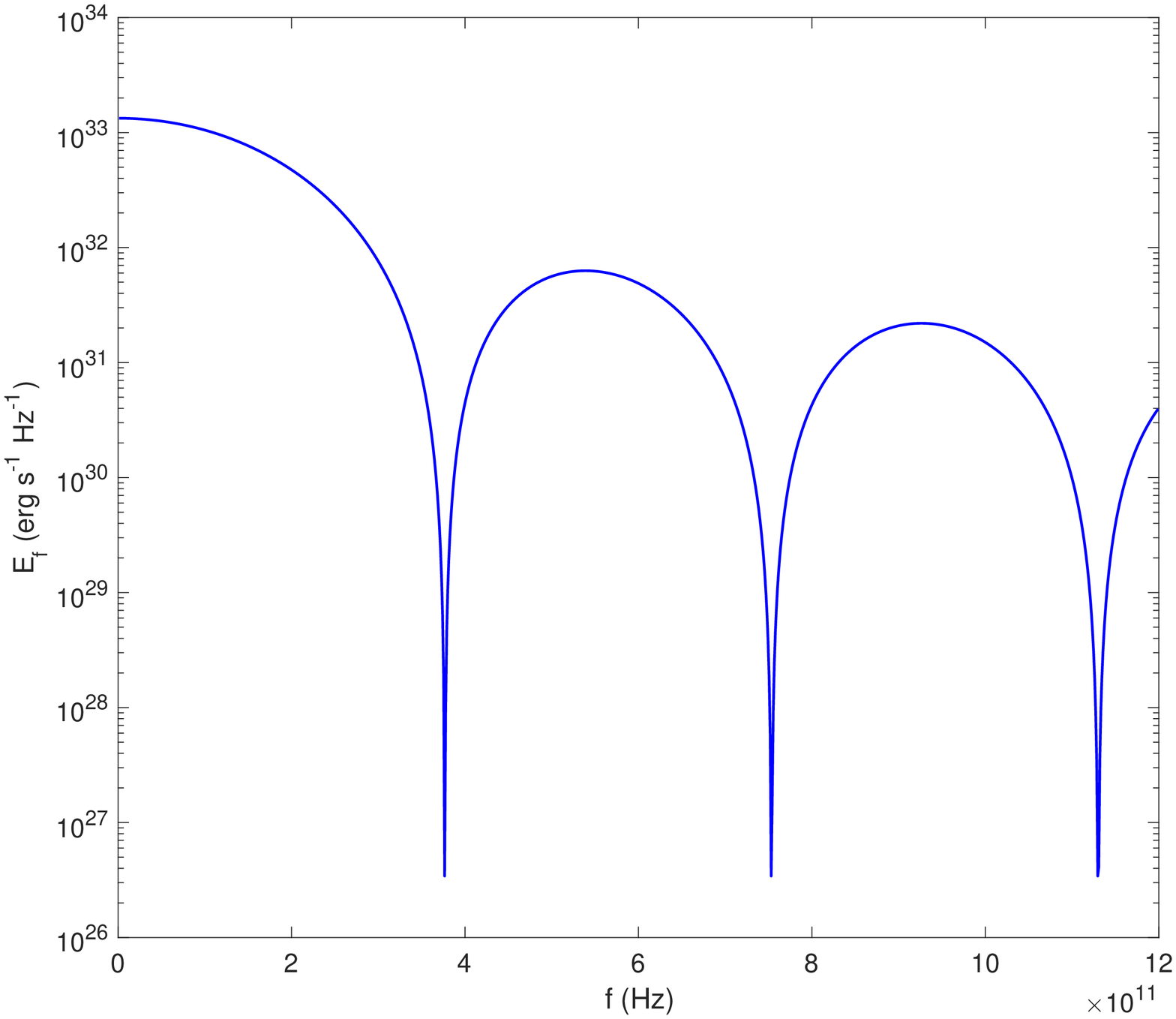}}
  \caption{The dependence of $E_f$ on frequency. The set of parameters is the same as in the previous graph except $n = 1 cm^{-3}$ and $D = 50000$ pc.}\label{fig1}
\end{figure}

\section[]{Main considerations}

As we have already shown in the previous work, the mass rate for a spherical probe is given by \citep{probe1}
\begin{equation}
\label{mt} 
\frac{dM}{dt} = \pi\beta r^2m_0n c,
\end{equation}
where $r$ is the radius of the probe, $\beta=\upsilon/c$ is a dimensionless velocity, $c$ represents the speed of light and $m_0$ is the mass of a molecule the probes encounter and $n$ is the number density of particles in a cloud. By taking into account an expression for the mass of the spherical probe, $M = 4\pi\xi r^3\rho/3$, the timescale of replication writes as 
where $r$ is the radius of the probe, $\beta=\upsilon/c$ is a dimensionless velocity, $c$ represents the speed of light and $m_0$ is the mass of a molecule the probes encounter and $n$ is the number density of particles in a cloud. By taking into account an expression for the mass of the spherical probe, $M = 4\pi\xi r^3\rho/3$, the timescale of replication writes as

$$\tau = \frac{M}{dM/dt} = \frac{4\xi}{\beta}\times\frac{\rho}{m_0n}\times\frac{r}{c}\simeq 3.4\times \frac{\xi}{0.1}\times$$
\begin{equation}
\label{tau} 
\times \frac{0.01}{\beta}\times\frac{\rho}{0.4g\;cm^{-3}}\times\frac{10^4cm^{-3}}{n}\times\frac{r}{0.1 mm} \;yrs,
\end{equation}
where $\xi$ denotes the fraction of the total volume filled with the material the probe is made of $\rho$ is the corresponding density normalised by the Graphene density (up to now the strongest material). As it is clear from the above expression time required for replication of a single probe is of the order of several years. This in turn means that in a nebula with a length-scale $2$pc the multiplication factor, $2^{t/\tau}$, becomes of the order $10^{48}$, which indicates high efficiency of the process. By assuming the initial number of robots $100$, one can straightforwardly show that an average distance between probes is of the order of $1$m. The time-scale has been estimated for the typical parameters of HII interstellar atomic hydrogen cloud \citep{carroll}.

In paper-I we have discussed that the probes when collecting the cloud atoms or molecules will inevitably lead to observable events. In particular, it is easy to show that a single probe capturing protons will accelerate them with the acceleration, $a\sim \upsilon^2/(2\kappa r)$ ($\kappa\leq 1$ is a dimensionless parameter characterizing an acceleration distance). This process in turn will lead to the Larmor emission and as a result, by means of the multiple events of proton capture, the bolometric luminosity corresponding to a single probe is given by \citep{probe1}
\begin{equation}
\label{lump} 
L_p\simeq\frac{\pi}{3\kappa} nrce^2\beta^4,
\end{equation}
leading to the total luminosity of the ensemble of self-replicators
\begin{equation}
\label{Lt1} 
L_{tot}\simeq\frac{\pi}{3\kappa} nrce^2\beta^4N_0\times 2^{t/\tau},
\end{equation}
where $N_0$ indicates the initial number of robots. Throughout the paper we use $\kappa = 0.1$.

As it has been shown by \cite{probe1}, for a given medium there is a certain relation between the probe size and the length-scale of a space, $D$, the ensemble is supposed to occupy reaching some critical (maximum) luminosity, $L_{cr}$

\begin{equation}
\label{rmin} 
\frac{3{\ln2}}{4\xi}\times\frac{m_0n}{\rho}\times\frac{D}{r}\simeq\ln\left(\frac{L_{cr}}{L_0(r)}\right),
\end{equation}
where $L_0 = N_0L_p$ is the initial total luminosity. Unlike the previous study, in this paper we consider both types of advanced civilizations: Type-II, harnessing the total energy of its host star and Type-III utilising the energy of a galaxy \citep{dyson}. For Type-II probes the critical luminosity is of the order of the Solar luminosity, $L_{\odot}\simeq 3.8\times 10^{33}$ergs s$^{-1}$, whereas for Type-III the critical value should be of the order of the bolometric luminosity of the galaxy. As an example we examine the Milky Way galaxy $L_{cr}\simeq 3.6\times 10^{10} L_{\odot}$ \citep{carroll}. The aforementioned equation should be solved numerically and one can show that as for a spherical molecular cloud with a length-scale $D\simeq 2$pc one obtains the probe size of the order of $0.013$cm and for the latter, by taking into account the diameter of the Milky Way galaxy $D\simeq 50000$pc, one has $0.025$cm. The calculations have been performed for $\beta = 0.01$.

In the previous study we have considered only a general emission pattern, without going into special characteristics. On the other hand, identification of the probes requires a detailed study of of the emission process. Therefore, the major aim of the paper is to examine the spectral energy distribution of self-replicating Von-Neumann probes. For this purpose we consider radiated energy per unit of frequency \citep{rybicki}
\begin{equation}
\label{W1} 
\frac{dW}{d\omega} = \frac{8\pi\omega^4}{3c^3}\times|\hat{d}(\omega)|^2,
\end{equation}
where $\omega$ is the cyclic frequency of emission and
\begin{equation}
\label{d1} 
\hat{d}(\omega) = \frac{1}{2\pi}\int_{-\infty}^{+\infty} d(t) e^{i\omega t}d\omega
\end{equation}
is the Fourier component of the dipole moment. By considering the Fourier transform of its second time derivative, one obtains \citep{rybicki}
\begin{equation}
\label{d2} 
-\omega^2\hat{d}(\omega) = \frac{e}{2\pi}\int_{-\infty}^{+\infty} a(t) e^{i\omega t}d\omega,
\end{equation}
where $e$ is the proton's charge and $a(t)$ is the corresponding acceleration when encountering the probes. If one approximates it as a constant value, $a$, during an interaction time $\tau_0\simeq 2\kappa r/v$, then it is straightforward to show
\begin{equation}
\label{d3} 
\hat{d}(\omega) = \frac{ea}{\pi\omega^3}\left|\sin\left(\frac{\omega\tau_0}{2}\right)\right|,
\end{equation}
reducing Eq. (\ref{W1}) to the following form
\begin{equation}
\label{W2} 
\frac{dW}{df} = \frac{4\pi e^2\beta^2}{3c}\times\left(\frac{f_0}{f}\right)^2\sin^2\left(\frac{f}{f_0}\right),
\end{equation}
where $f_0\equiv\beta c/(2\pi \kappa r)$ and we have taken into account $\omega\equiv 2\pi f$. The derived formula characterises a single event of a proton capture. To obtain the total emission per unit time and unit frequency interval (spectral power), $E_f \equiv dW/(dtdf)$, we note that the incident proton flux is $n_pc\beta$, then one can write
\begin{equation}
\label{W2} 
\frac{dW}{dtdf} = n_p N(t)\frac{4\pi e^2r^2\beta^3}{3}\times\left(\frac{f_0}{f}\right)^2\sin^2\left(\frac{f}{f_0}\right),
\end{equation}
where 
\begin{equation}
\label{N} 
N(t) = N_0\times e^{t/\tau},
\end{equation}
is the total number of probes. As a next step, we are going to apply the obtained expressions to Type-II and Type-III civilizations.

\subsection{Type-II interstellar probes}

In this subsection we examine the Von-Neumann probes belonging to Type-II societies. For this purpose we consider a spherical nebula with a radius of the order of $1$pc, then, as we have already shown the optimal length-scale of the robot should be of the order of $0.013$ cm leading to the corresponding normalizing emission frequency \footnote{The velocities of the probes are not relativistic and therefore, the Doppler shift does not affect the result.}  
\begin{equation}
\label{freq} 
f_0\simeq\frac{\beta c}{2\pi r\kappa}\simeq  3.7\times 10^{10}\times\frac{\beta}{0.01}\; Hz.
\end{equation}

In Fig. 1 we plot the dependence of $E_f$ on frequency. The set of parameters is: $\beta = 0.01$, $N_0 = 100$, $\xi = 0.1$, $m_0 = m_p$ ($m_p$ is the proton mass), $\rho = 0.4g\;cm^{-3}$, $n = 10^4cm^{-3}$, $D = 2$ pc. As it is clear from the figure, the spectral power of Von-Neumann probes has a specific fingerprint: at certain frequencies $E_f$ diminishes. From Eq. (\ref{W2}) one can straightforwardly show that the spectral power becomes zero for the following discreet set of equally spaced frequencies $f_k = f_0\pi k$ for $k=1,2,...$. The presented spectral behaviour is drastically different from black body spectrum as well as from the thermal bremstrahlung, which can be a very important factor in identifying the Von-Neumann interstellar probes. Another interesting feature is that if one considers a different value of $\beta$, the corresponding length-scale of the probe will be different, but the overall spectral feature will be the same - in this sense the given spectrum is unique. The initial number of probes $100$, very rapidly multiplies and approximately in $650$ yrs the total number will be of the order of $10^{50}$ colonising the whole nebula. One should also note that if one maintains an annual growth rate of $1 \%$ in industry, then $3000$ yrs is quite enough to reach such an advanced technological level, which means that the assumption of existence of such alien societies is quite reasonable \citep{dyson}.

\subsection{Type-III interstellar probes}

As it has been classified by \cite{kardashev} advanced civilizations consuming almost the whole power of their host galaxy ($\sim 4\times 10^{44}$ergs s$^{-1}$) is the so-called Type-III society. Our own civilization consumes approximately $4\times 10^{19}$ergs s$^{-1}$, which means that if an average $1\%$ of annual growth is maintained our society will achieve Type-III in $\sim 6000$ yrs, which like the previous case, is a negligible value on the galactic time-scale.

For making estimates, we consider the parameters of the Milky Way, having the diameter $50000$ pc and the total luminosity $\sim 3.6\times 10^{10}L_{\odot}$. Then, by taking into account the number density of interstellar protons  $\sim 1$ cm$^{-3}$ for the optimal size of the probe one obtains $0.025$cm (see Eq. (\ref{rmin})), which means that the frequency describing the spectral energy distribution is of the order of $1.9\times 10^{10}$ Hz.

In Fig. 2 we show the behaviour of $E_f$ versus $f$. The set of parameters is the same as for Type-II civilizations except $n = 1$ cm$^{-3}$ and $D = 50000$ pc. Likewise the previous case, one can see that emission spectra of Type-III probes vanishes for a discreet set of frequencies $f_k = f_0\pi k$ for $k=1,2,...$ with $f_0 = 1.9\times 10^{10}$ Hz. For the considered value of velocity, the time-scale of colonising the whole galaxy is of the order of $T\simeq D/(2\beta c)\simeq 16$ Myr, which is by three orders of magnitude less than the galactic age, $10$ Gyr, therefore, colonization in the context of time-scales might be quite reasonable. From Eq. (\ref{N}) one can straightforwardly show that after the mentioned time, the multiplication factor will be of the order of $10^{58}$ leading to the total luminosity of the order of the galactic luminosity. Peculiarity of the emission spectra, characterised by equally spaced dips and the overall dependence on frequency, which is different from all other spectral behaviours might be a good sign to identify the Type-III interstellar Von-Neumann probes.

Recently new prospects appeared thanks to the Five-hundred-meter Aperture Spherical radio Telescope (FAST)  - a single-dish radio telescope, which is going to work in the SETI projects as well \citep{fast}. On the other hand, radio emission of Von-Neumann probes is significant. In particular, for Type-II probes the total power emitted in the Hydrogen line ($1.42$ GHz) will be of the order of $10^{32}$ ergs s$^{-1}$, therefore FAST might potentially detect such objects.

\section{Conclusion}

We have studied the spectral characteristics of Type-II and Type-III Von-Neumann self-reproducing interstellar probes. For this purpose we have considered emission generated by proton capture and examined the corresponding mechanism of bremstrahlung.

Considering Type-II robots it has been shown that to colonise the spherical nebula with radius of the order of $1$ pc, the optimal length-scale of the probe should be $0.013$ cm, leading to emission from radio to infrared. It has been found that the differential emission power has equally spaced dips in frequency. The same behaviour has been found for Type-III interstellar probes as well (also potentially observable from radio to infrared), which might be a significant indicator to identify these exotic robots.

Generally speaking, we have analysed the emission pattern for certain typical parameters. Consequently for different values the emission spectra will be different, but what is very important, a general fingerprint - the particular behaviour, with periodic dips - is maintained and in this sense the mentioned results are unique. Therefore, if one finds the similar dependence of spectral power on frequency, this might be a significant result indicating the existence of Von-Neumann interstellar probes.


\section*{Acknowledgments}
The research was supported by the Shota Rustaveli National Science Foundation grant (NFR17-587).

\end{document}